\newcommand{\beq}{\begin{quote}}
\newcommand{\enq}{\end{quote}}
\newcommand{\be}{\begin{equation}}
\newcommand{\en}{\end{equation}}
\newcommand{\del}{\delta}
\newcommand{\om}{\omega}

\newcommand{\De}{\Delta}

\newcommand{\eps}{\epsilon}
\documentclass[preprint2]{aastex}
\begin{document}
\title{ Stability and eccentricity of
periodic orbits for two planets in a
1:1 resonance. 
} 
\author{Michael Nauenberg}
\affil{Physics Department, University of California,
    Santa Cruz, CA 95064}
\begin{abstract}
The nonlinear stability domain of Lagrange's celebrated 1772  solution of 
a three-body problem is obtained numerically as a function of the  masses
of the bodies and the common  eccentricity of their Keplerian  orbits. 
This domain shows that this solution  may be realized in extra-solar
planetary systems similar to those that have been discovered recently
with two Jupiter-size  planets orbiting a solar-size  star.
For an exact 1:1 resonance, the  Doppler shift variation in
the emitted light  would be the same  
as for stars which have only a single planetary companion. 
But it is more likely that in actual extra-solar planetary systems
there are deviations from such a resonance, raising the interesting prospect 
that Lagrange's solution  can be identified by an analysis of the observations.
The existence of another stable  1:1 resonance solution 
which would have  a more unambiguous Doppler shift  signature is also  discussed. 
\end{abstract}

\section{Introduction}
In many extra-solar planetary systems discovered recently,
the observed Doppler shift of the emitted  light
is well described by assuming that the central
star is moving in a  Keplerian elliptic  orbit 
due to its gravitational interaction with a single Jupiter-size planet.
If residuals  are present in a least squares fit to
the data after possible chromospheric  fluctuations in the star
have been taken into account, these residuals signal the presence 
of an additional planet, or possibly several planets 
\citep{marcy0} \citep {marcy1}. There is, however, an important exception
when such a star also travels  on a  Keplerian orbit even
though there are {\it two}  planets orbiting it. 
We have in mind  Lagrange's celebrated  solution of the
three-body  problem \citep {lagrange}, 
for which he won the prize of the Royal Academy of Science of
Paris. In this solution,  each body moves 
on a Keplerian  orbit with a common plane, period, eccentricity 
and focus that is located at their center of mass, in such a manner 
that at all times  the relative positions of these bodies  
form the  vertices of an equilateral triangle of variable size, (see
Fig. 1). This solution will be discussed in section 2.
It is therefore interesting to consider the possible  occurrence 
of such a 1:1 resonance in
extra-solar planetary systemm with
two Jupiter-size planets,  particularly
in view of the recent  discovery of a remarkable  2:1 resonance
of two such large planets in GJ876 \citep {marcy2}.
To be relevant to current astronomical discoveries, however,
it is necessary that Lagrange's solution be stable
in the range of observed masses and eccentricities.  
In the past, linear stability analyses have been carried out that were 
primarily focused  on the restricted three-body problem (where
one of the three masses vanishes), which is applicable to the study
of the motion of  asteroids, but these results can also be extended to the general
Lagrange solution as will be discussed in  section 3.
In this paper  we present  the results of a  numerical analysis of the 
nonlinear stability domain of Lagrange's solution as a function of the 
masses of the three bodies and the eccentricity of the common 
elliptical orbits (see  section 4). In practice  there will be 
deviations from Lagrange's solution, and the interesting  question 
arises as to whether there are distinct characteristics 
which would distinguish these deviations  from other types
of perturbations due to additional planets. While 
variations in the eccentricity and major axis of the approximate
elliptical orbit are common to all  perturbations caused by 
the presence of a second planet, one of
the most  distinguishing feature of a slightly off-resonance but
stable  Lagrange solution is that the
the  rotation rate  of the axis of the elliptic orbit 
of the star is much smaller than for other types of perturbations, (see Fig. 2).

\begin{figure}
\plotone{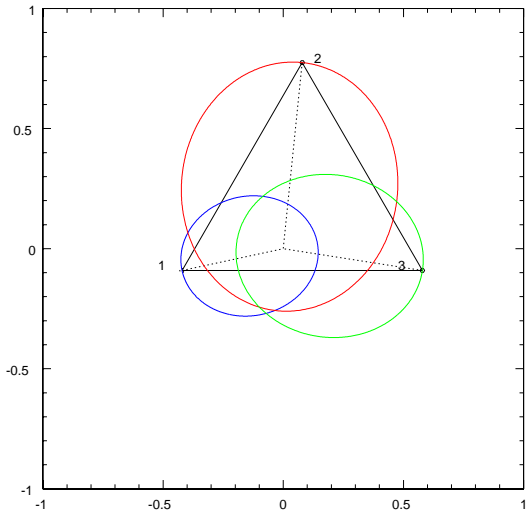}
\caption{ Lagrange's periodic 1:1 resonance  solution for the three-body problem,
showing the locations of the bodies at apogee on the vertices of 
an equilateral triangle (full lines). The dashed lines show
the direction of the major axis for each ellipse,  and their
intersection at the center of mass, which is their  common focus. 
}
\end{figure}

\begin{figure}
\plotone{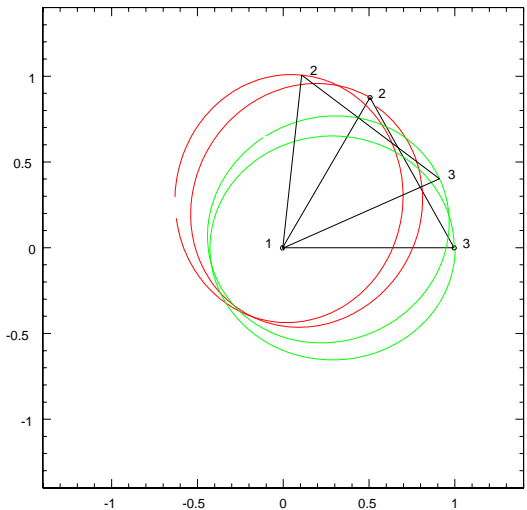}
\caption{ Rotation of the axes of the  ellipses in Lagrange's
solution after $800$ periods, for a one percent deviation in the initial
position of planet 2 from exact resonance. 
}
\end{figure}
For example, for a $1\%$ deviation in the position of the lighter planet,
assumed to be about a Jupiter mass, we find that
it takes about  $800$ periods for a  complete  revolution of the major axis,
(see Fig. 2).
In contrast, a fit to  the recently discovered 2:1 resonance
in GJ876 indicates that the major axis of the planets  
should complete a  revolution  in about $53$ periods of the heavier planet
\citep {michael2}.   

Lagrange's solution is not the only 1:1 resonance for the three-body
problem. We also  consider  another solution  that we found to be
stable in the domain of masses and eccentricities relevant to extra-solar
planetary systems  \citep {laughlin}.  In this case the two lighter bodies 
(planets),  and the heaviest body (star) 
have different  orbits, and therefore the Doppler shift data should be   
readily distinguishable from the case of a single planet.
The characteristic feature of this solution is that at each half period the 
three bodies are aligned,  and if the heavier planet
is in a nearly circular orbit, the lighter planet moves in a highly
eccentric  elliptical orbit, (see Fig. 3).
A periodic   alignment of the two planets and the star  is also a  characteristic of 
other resonance solutions, as in the case of the 2:1 resonance in
GJ876\citep{laugh2} \citep{rivera} \citep{peale}  \citep{michael2}. The evolution
of this configuration when the planets are slightly off resonance is shown
in Fig. 4, which illustrates the clockwise  rotation of the major axis of the
eccentric orbit.

\begin{figure}
\plotone{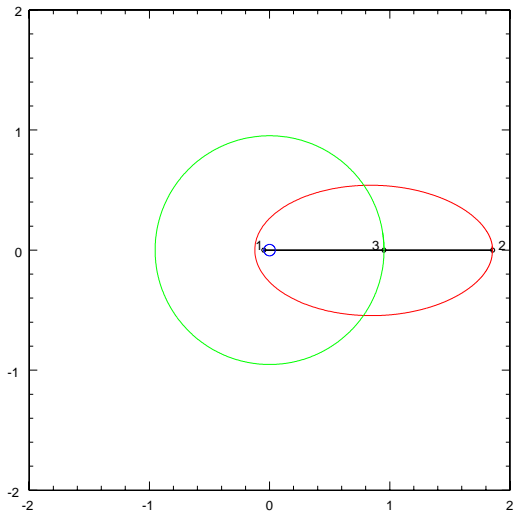}
\caption{ Another  1:1 resonance  solution for the three-body problem,
showing the inner heavier planet (green) on a 
nearly circular orbit, the lighter planet (red) on
an eccentric elliptic orbit with eccentricity $\eps =.8$,
and the central star (blue), as they
appear aligned at maximum elongation of the ellipse.
}
\end{figure}

\begin{figure}
\plotone{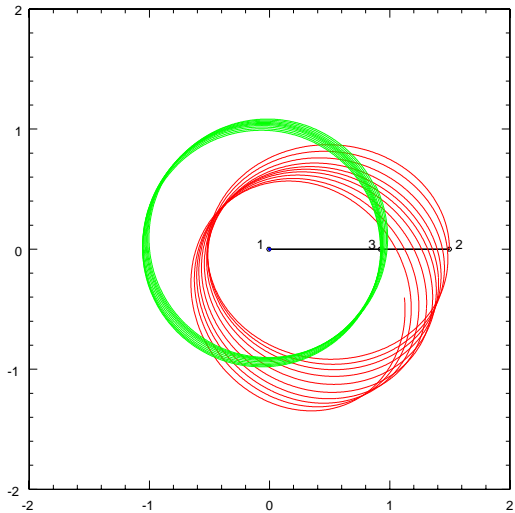}
\caption{ Evolution of the orbits shown in Fig. 2
for a small deviation from 1:1 resonance, showing the 
rotation of the major axes of the planets
during 10 periods.
}
\end{figure}

\section{ Lagrange's solution of a three-body problem}

In Lagrange's solution \citep{lagrange} of a three-body problem,
each body travels on a separate  elliptic orbit 
with a common  period, eccentricity  and focus which is located at
their center of mass, in such a way that these bodies 
are always at the vertices of an equilateral 
triangle of variable size. An example is  illustrated in Fig.1, 
where the  masses are  in the ratio $1.0:0.7:0.2$, and the common eccentricity
is $\eps=.5$,  showing the equilateral triangle for the relative positions 
of the three bodies  at apogee. 
Apart from the overall orientation of the system,
these parameters  uniquely describe the solution.
In this solution the relation  
between   variables is  somewhat different 
from the corresponding  ones in the two-body problem. 
For example, the common frequency $\om$ 
or period $P=2\pi/\om$ of the motion is  given by
the relation
\be
\label{kepler1}
\om=\sqrt{Gm (1+\eps)^3/R^3},
\en
where $G$ is Newton's constant, $m=m_1+m_2+m_3$ is the total mass, 
$m_1$ is the mass of the heaviest body (star), $m_2$, $m_3$ 
are the masses  of the lighter bodies (planets),
$\epsilon$ is the common eccentricity, and $R$ is the  maximum size
of the equilateral triangle on which the three bodies are located.
Then the major axis $a_i$ of each of the elliptic orbits  is 
\be
a_i=\frac{\sqrt{m_j^2+m_k^2+m_jm_k}}{m(1+\eps)}R,
\en
where the subscripts $ i,j$ and $k$ are permutations of the
the integers $1,2,3$,
while  the  mean of the  velocity  of the star 
at the maximum and minimum distance from the foci of the ellipse is
\be
\label{vel1}
K=(2\pi G/P)^{1/3}\frac{\sqrt{m_2^2+m_3^2+m_2m_3}}{m^{2/3}\sqrt{1-\eps^2}}.
\en
If the planetary  masses $m_2$ and $m_3$ are small  compared to the mass
of the star, $m1$,  then $a_2 \approx a_3\approx a$, where $a= R/(1+\eps)$ and
$\om \approx \sqrt{Gm/a^3}$ as in the corresponding two-body problem. 
In principle, any  fit to the  data with a single planet of mass $m_p$
can also be attributed  to two planets which, according to 
Eq. \ref{vel1} for $K$, have masses satisfying the
relation  $m_p=\sqrt{m_2^2+m_3^2+m_2m_3}$, 
which is somewhat less than the sum of the masses of the two planets.
In practice, however, it is unlikely that extra-solar planetary
systems would  occur in an exact 1:1 resonance, and therefore the presence
of a second planet manifests itself  in the occurrence of 
residuals in a single Keplerian orbit fit to the data.

\section {Linear stability analysis}
The first linear stability analysis of Lagrange's  solution
was carried out by Routh 
for the special case of circular orbits \citep{routh}
\footnote {In this paper, Routh incorrectly  attributed  Lagrange's solution 
to Laplace.}. 
Assuming that the attractive
forces between the bodies depends on the relative distance $r$ 
as $1/r^{\kappa}$, Routh
demonstrated that Lagrange's
solution was stable provided that the masses satisfied the inequality

\be
\label{stab1}
\gamma <\frac{1}{3}(\frac{3-\kappa}{1+\kappa})^2,
\en
where 
\be
\label{gamma1}
\gamma= \frac{(m_1m_2+m_1m_3+m_2m_3)}{(m_1+m_2+m_3)^2}.
\en
For gravitational interactions where  $\kappa=2$, the constant on the right 
hand side 
\footnote { Routh found
that he had been anticipated in this important result
by someone called  M. Gascheau in a thesis on mechanics,
apparently left unpublished }
of Eq. \ref{stab1} is $1/27$, 
which  implies that the masses of the 
two lighter bodies (planets),  $m_2$ and $m_3$,
must be much smaller than the mass of the heaviest body (star) $m_1$.
For example, setting  $m_2=0$, which correspond to the restricted 
three-body problem and applies to the motion of asteroids such as the Trojans, 
this inequality implies that
$m_3/(m_1+m_3) <.03852..$  which has become known as Routh's critical
point. Neglecting quadratic terms in the mass ratios $m_2/m_1$ and
$m_3/m_1$, Routh's inequality  becomes approximately 
$(m_2+m_3)/m_1<1/27$, indicating that the stability depends to a very
good approximation only on the sum of the masses of the lighter bodies. 

For the stable configurations, Routh obtained  the frequencies $\om_1,\om_2$ 
and $ \om_3$ of the normal modes in the plane of the orbits, 
which for $\kappa=2$ are given by 

\be
\label{om1}
\om_1=\om \sqrt{\frac{1}{2}(1+\sqrt{1-27\gamma)}},
\en

\be
\om_2=\om \sqrt{\frac{1}{2}(1-\sqrt{1-27\gamma)}},
\en

\be
\label{om3}
\om_3=\om,
\en
where $\om$ is the fundamental Kepler frequency, Eq. \ref {kepler1},
and he determined the corresponding amplitudes for these  modes. 
While Routh did not discussed the stability with respect to deviations
perpendicular to the plane of the orbit, it is straightforward to show that
for this case the  Lagrange  orbits are stable.
Remarkably , he  also briefly  considered second order deviations, 
which he remarked could ``ultimately disturb the stability'', 
but this part of his analysis was  incomplete, calling 
attention only to two of the possible  commensurability or resonance relations,
$\om_3=2\om_2$ and $\om_1=2\om_2$, for which $\gamma=1/36$ and $\gamma=16/675$,
respectively.   

For elliptic orbits, a linear stability analysis of Lagrange's solution was 
not  carried out until some 90 years later, when  \citep{danby} numerically
integrated  the Floquet equations for the first
order deviations from the Lagrange solution 
of the  restricted three - body problem (see also \citep {bennett}. 
Subsequently,  a majority of  
stability  studies have been confined to this special case,  
but as we shall see, the results can
also be applied to the general solution \citep{marchal}.
For a modern discussion of 
the stability of the Lagrange solution see \citep {siegel}

\section{Nonlinear stability domain of Laplace's solution}

The nonlinear stability domain of Lagrange's  solution for the three-body
problem that is  presented here was obtained  by integrating  the equations
of motion numerically, and determining whether,
for a small initial deviation from the solution, the orbits were 
either confined 
or unconfined after a large number $n$ of periods.
We found that when  $n$ was increased  from 400 to 800, 
there were no significant changes in the results   
except near the critical points discussed below, where we increased
$n$ until no further changes occurred.
For the deviations we consider small displacements of the 
position of one of the lighter bodies {planets} in the plane of the orbit and  
also perpendicular to this plane. As we shall see, except in the case when $\om_3=2\om_2$,
only initial  deviations in the plane of the orbit gave evidence for some of the expected  nonlinear 
instabilities  near the commensurability relations for the three fundamental  
frequencies,  Eqs. \ref{om1}-\ref{om3}.

Starting with a small deviation in the velocity of one
of the lighter bodies of order $1 \%$ in the direction perpendicular 
to the plane of the orbits,
our numerical result for the nonlinear stability domain is shown in Fig. 5, 
where the unstable regions are indicated  by small squares.  
In agreement with  the linear stability analysis ( see section 3),
the nonlinear stability  domain  depends 
only on the common eccentricity $\eps$  of the orbits (horizontal axis),
and  the Routh parameter $\gamma$, Eq. \ref{gamma1} (vertical axis). 
This parameter  is determined by
the ratios $m_2/m_1$ and $m_3/m_1$  of the masses of the 
lighter bodies (planets) 
to the mass $m_1$ of the heaviest body (star) (vertical axis). 
We have verified this result  by  computing this domain  for different 
fixed values for the ratio of the
masses of the planets, without observing any changes when plotting
the results with  Routh's parameter $\gamma$. 
For the particular case shown in Fig. 5, the computation was  carried out for
$m_2=m_3$, while the curves shown are the linear stability
computations of Danby and Bennett \citep{danby} \citep{bennett},
which were  originally obtained  for the restricted three-body problem
where either $m_2$ or $m_3$ is set  equal to zero.

\begin{figure}
\plotone{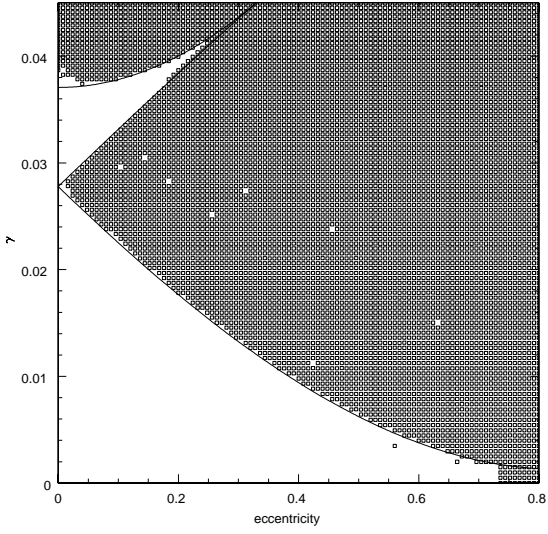}
\caption{ Stability domain for Lagrange's solution as a function
of the eccentricity $\eps$ and the Routh parameter $\gamma$, Eq. \ref{gamma1}, obtained
for initial deviations perpendicular to the plane of the orbit.
}
\end{figure}

This nonlinear stability domain  looks surprisingly similar 
to the stability domain 
of the  Mathieu equation \citep{mathieu} with an additional
nonlinear restoring or damping
term. Indeed, for the restricted three-body problem,
a linear stability analysis of the  equations of motion, 
to first order in the common eccentricity $\epsilon$, in a frame of
reference rotating with the frequency of the orbits, yields 
a bifurcation at  $\om_3=2\om_2$  which can be viewed as a 
parametric resonance between the fundamental orbital 
frequency for elliptic motion
and the oscillation  frequency of the  first order  deviations of the massless
body. In the general case, we have seen that this bifurcation occurs for $\eps=0$
at $\gamma=1/36 = .02777...$. For the restricted problem, $m_2=0$, 
this  corresponds to   $\De_0 =2\sqrt2/3\approx .9428$, 
where $\De=(m_1-m_3)/(m_1+m_3)$  and  $m_3/m_1 \approx .02944$. 
We evaluated the instability domain to first order in the
eccentricity $\eps$, and found that it lies
inside the wedge  
\be
\label{slope1}
\De_0-\del\epsilon \leq \De \leq \De_0+\del\epsilon, 
\en
where  $\del=3/20\sqrt2\approx .10606$. 
This domain is shown in  Fig. 6, together with  the corresponding 
nonlinear domain  where the orbit actually becomes unstable. 
These linear boundaries mark the onset of
a bifurcation in which the orbits first  become aperiodic  filling
out a confined region of space.

\begin{figure}
\plotone{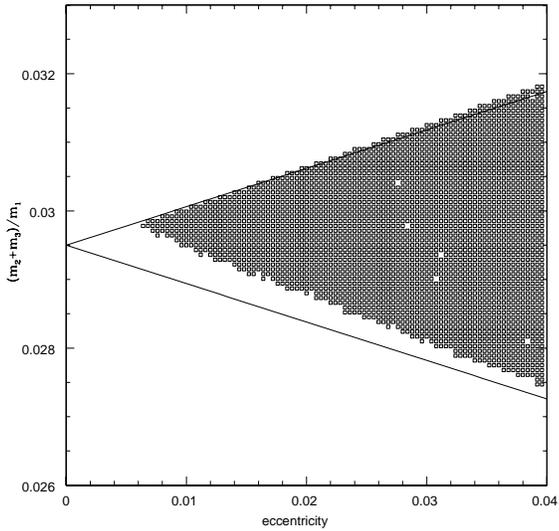}
\caption{ Enlargement of the stability domain shown in Fig. 5 
near the critical point for the parametric resonance at $\om_3=2\om_2$.
}
\end{figure}

An enlargement of the stability domain  near the critical point at $\eps=0$ and
$\gamma=1/27 \approx .037037..$ is shown in Fig. 7, which displays our
results for the case that $m_2=m_3$. 
This critical point is often  discussed in connection with the stability 
of the Lagrangian points for the restricted three body problem 
in a frame of reference rotating with the frequency of the 
circular orbits, and is  applied to the study of the Trojan 
and other asteroids \citep {murray}.
As in the previous case, however, this critical point marks only a bifurcation
to an aperiodic but confined motion, while numerically 
we find that  the nonlinear instability begins at a somewhat higher value
of $\gamma \approx .0391$, corresponding for $m_2=0$ to  $m_3/m_1
\approx .0425$. A  curve quadratic  in the eccentricity fits
well  the boundary of the upper part of the nonlinear instability domain,
as shown in Fig. 7, while
a linear curve  fits the lower part of the  domain 
with  a slope which is somewhat higher than the one
which we calculated analytically in the linear approximation, Eq. \ref{slope1}
but in good agreement
with the numerical linear stability results of \citep{danby} and \citep{bennett}.

\begin{figure}
\plotone{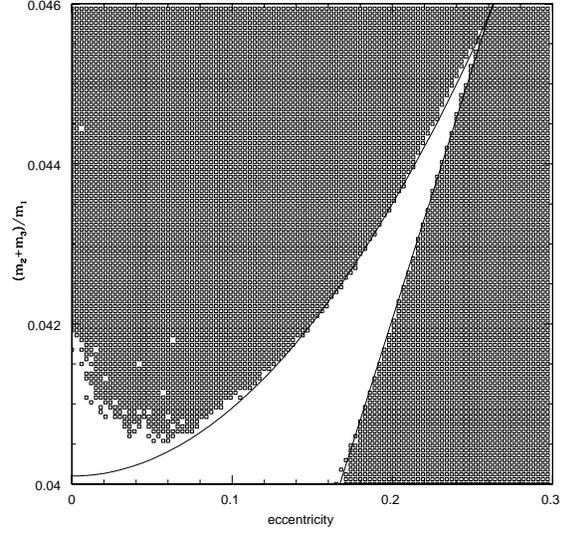}
\caption{ Enlargement of the  stability domain shown in Fig. 5  
near the critical point at $\om_1=\om_2$.  
}
\end{figure}

Up to now, we have considered only  initial deviations  in 
the direction perpendicular to the plane of Lagrange's orbits.
In Fig. 8 we show the  nonlinear stability domain obtained by taking
an initial a displacement $dx=dy=.001$ in this plane  for  one of 
the two  lighter bodies. This result was
obtained by fixing the ratio  $m_2/m_3=1$, but the same results 
are obtained for other 
values of the ratios of these masses
when the results are plotted as a function of the Routh parameter
$\gamma$, Eq. \ref{gamma1}. 
We see evidence in this figure for the  
instabilities at  the nonlinear  resonances  $\om_1=2\om_2$  and $\om_3=3\om_2$,
corresponding to $\gamma=16/675 \approx .02370.. $ and $\gamma=32/2187\approx .01463..$
Also, the domain of stability shrinks in the upper wedge 
with indications of additional resonances in this region. This is confirmed
by an  enlargement of this region, which 
shows the tail of the instability due to the
resonance at $3\om_1=4\om_2$ corresponding to 
$\gamma=576/16875 \approx .03413...$ as well as  other
resonances (see Fig. 9).

\begin{figure}
\plotone{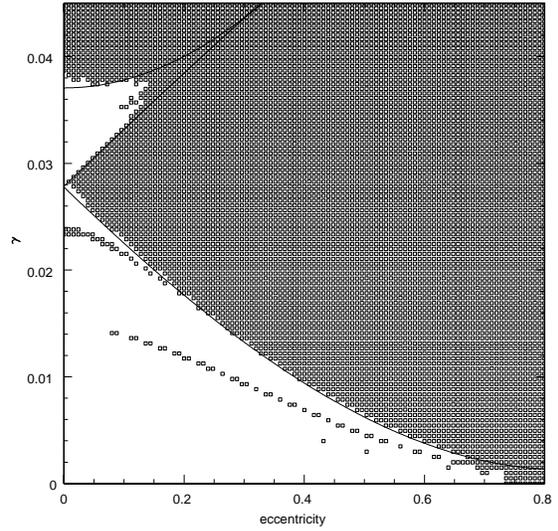}
\caption{ Stability domain for Lagrange's solution as a function
of the eccentricity $\eps$ and the Routh parameter $\gamma$, 
Eq. \ref{gamma1}, obtained for initial deviations in the plane of the orbit.
}
\end{figure}

\begin{figure}
\plotone{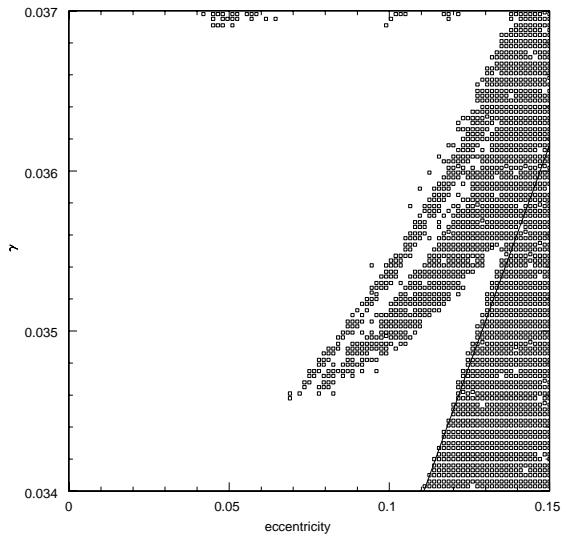}
\caption{ Enlargement of the stability domain for Lagrange's solution 
in the upper wedge shown in Fig. 8 as a function
of the eccentricity $\eps$ and the Routh parameter $\gamma$, 
Eq. \ref{gamma1}, obtained for initial deviations in the plane of the orbit.
}
\end{figure}

\section{ Another  1:1 resonance solution}
Lagrange's solution is not the only stable 1:1 resonance system for
the three-body problem. Another  solution is illustrated in Fig. 3, 
in which the two lighter bodies or planets are traveling  on two different
orbits which are approximately  elliptic in such a way that
the central star and the planets are aligned when located  at  the
maximum or minimum distance from the center of mass.
We found these periodic orbits \citep{laughlin} 
by a new method based on an  expansion 
of the coordinates in a Fourier series  with a common period, 
and we  determined  the Fourier coefficients of this expansion 
by finding  minima of the action integral with respect
to these coefficients by an iterative process \citep {michael1}.
A characteristic feature
of these orbits is that when  the heavier of the two planets is in a 
nearly circular orbit, the lighter planet is in a elliptical orbit 
with a very large  eccentricity 
which rises slowly with increasing 
mass of the heavier planet, as shown in Fig. 10 . 

\begin{figure}
\plotone{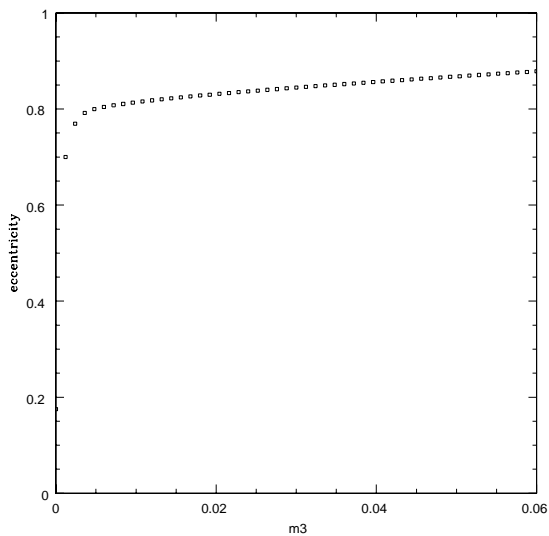}
\caption{ Dependence of the  eccentricity of the lighter planet on the 
mass of the heavier planet for the 1:1 resonance solution
shown in Fig. 3.
}
\end{figure}

In contrast to the Lagrange's  solution discussed previously, 
a small deviation from exact 1:1 resonance in this case 
leads to a relatively rapid rotation of the 
major axis of the elliptical orbit, as illustrated in  Fig 3.  
This leads to characteristic  modulations in the Doppler shift oscillations
of the light emitted by the star.
As in the case of Lagrange's solution discussed in the previous section,  
we found that these orbits are stable over the  range of masses 
relevant to extra-solar planetary systems, see Fig. 11.

\section{Conclusions}
The nonlinear stability domain for Lagrange's solution of the 
three body problem  
shown in  Fig. 8 indicates that there is a wide 
range of Jupiter-size  planetary  masses  (including 
brown dwarfs) and eccentricities for which  such solutions can exist in
extra-solar planetary systems.  
For example, for an eccentricity of $\eps \approx.6$ 
the ratio of the total
mass  of the two planets to the mass of the star for which the solutions 
are stable
is $.004$, except for a small region where  nonlinear resonances
occur. This mass  correspond to  $4.2$ Jupiter-masses
for a one solar-mass star, while for smaller eccentricities, $\eps \leq .2$,
there is a wedge of stable solutions for higher mass ratios up to approximately $.04$.
In principle, any  Doppler shift  data that can be fitted under the assumption
that  only a single planet is orbiting the central star, can  
equally well be attributed to two planets
orbiting the star according to  Lagrange's solution.
In practice, however, it is very  unlikely that two  planets are in 
an exact 1:1 resonance, and therefore  one expects to find 
residuals in the data  which signal the present of a  second planet.
One of the main effects due to the perturbations caused by a second planet
is the secular rotation of the major axes of the approximate Keplerian
ellipses that characterize the orbits of the planets and the star. 
This effect, however, would not occur in the case of an exact 1:1 
Lagrangian resonance, and it is strongly  suppressed in the case that
the resonance is  approximate.  It may be thought that in view of  
the greater number of degrees of freedom present with Lagrange's solution,
a better least squares  fit to the data 
should be readily available. This is not
the case, because Lagrange's equations also  
allow  for unphysical solutions where the mass of one or even  
both of the planets can have negative values, provided the
total mass is positive. Indeed, in
a preliminary attempt to obtain a least squares  
fit to  data which show residuals, the optimal  mass of 
the smaller of two  Lagrange  planets turned out to be negative,
which ruled out this solution.

The stability domain  for the type of 1:1 resonance solution
discussed in section 5 is shown in Fig. 11, demonstrating that this solution
also encompasses the possibility  of two
Jupiter-size planets orbiting a solar-size star. In this analysis  we 
restricted the heavier  planet to be in a nearly circular orbit,
and found that the lighter  planet is in a highly eccentric
orbit with $\eps \approx .8$ , (see Fig. 4). If this restriction
is  relaxed, we also find similar  stable solutions,
and for equal masses the  two planets  exchange eccentricity 
when the major axis rotates through $180$ degrees.

In summary, it is likely that extra-solar planetary systems 
that have several Jupiter-size
planets which are  close enough to give rise to  significant 
gravitational perturbations will be in resonance,
because numerical investigations have shown that such systems 
can be stable. In such cases, the planets and the central
star are periodically  aligned. An interesting  exception
is the  1:1 resonance solution of Lagrange, where the planets
and the star are located at all times on the vertices of an equilateral
triangle of varying size. It would be very  exciting if this solution,
discovered by Lagrange  230 years ago, and realized thus far only in the motion
of the Trojan  and other asteroids in our solar system
\citep {murray}, would be also present in the orbits of planets 
in extra-solar systems. Likewise a search should be undertaken
to  find also two planets in extra-solar systems which are  
in a 1:1 resonance of the type discussed in section 5, which 
does not occur in  our system.

\begin{figure}
\plotone{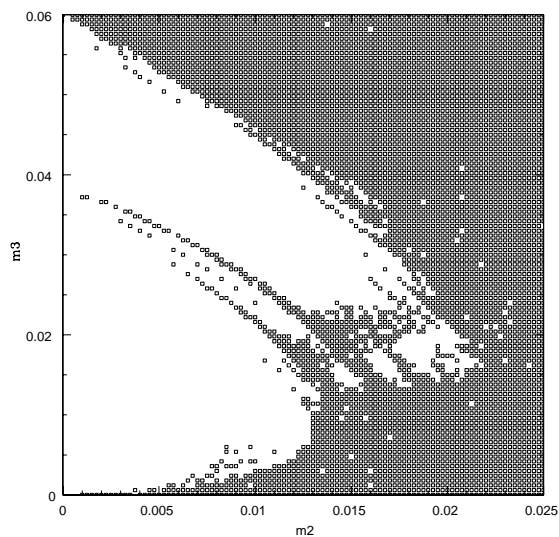}
\caption{ Stability domain for the  1:1 resonance configurations
of the type shown in Figs. 3 and 4.
}
\end{figure}

\acknowledgements
I would like to thank Richard Montgomery and Carles Simo for useful references to
the vast literature on the  Lagrange solution of a three-body problem.

\email{michael@mike.ucsc.edu}.

{}

\end{document}